\documentclass[fleqn,usenatbib]{mnras}

\usepackage{newtxtext,newtxmath}
\usepackage[T1]{fontenc}

\DeclareRobustCommand{\VAN}[3]{#2}
\let\VANthebibliography\thebibliography
\def\thebibliography{\DeclareRobustCommand{\VAN}[3]{##3}\VANthebibliography}

\usepackage{graphicx}	% Including figure files
\usepackage{amsmath}	% Advanced maths commands
\usepackage{amsfonts}
\usepackage{float}
\usepackage{subcaption} % Sub-figures and Sub-tables
\usepackage{siunitx}    % Some special units
\usepackage{wrapfig}
\usepackage{adjustbox}
\usepackage{multirow}

\title[SMBH parameter estimation with EHT observables]{EHT observables as a tool to estimate parameters of supermassive black holes}

\author[M. Afrin and S. Ghosh]{
Misba Afrin,$^{1}$\thanks{E-mail: me.misba@gmail.com}
and Sushant~G.~Ghosh$^{1,2}$
\\
$^{1}$Centre for Theoretical Physics, Jamia Millia Islamia, Maulana Mohammad Ali Jauhar Marg, New Delhi 110025, India\\
$^{2}$Astrophysics and Cosmology Research Unit, School of Mathematics, Statistics and Computer Science, University of KwaZulu-Natal, University Road\\ Westville, Private Bag 54001, Durban 4000, South Africa
}

\date{Accepted 2023 July 5. in original form 2023 June 19}

\pubyear{2023}

\begin{document}
\label{firstpage}
\pagerange{\pageref{firstpage}--\pageref{lastpage}}
\maketitle

% Abstract of the paper
\begin{abstract}
The Event Horizon Telescope (EHT) collaboration unveiled event-horizon-scale images of the supermassive black holes (SMBHs) M87* and Sgr A*, revealing a dark brightness depression, namely the black hole shadow, whose shape and size may encode the parameters of the SMBHs, and the shadow is consistent with that of a Kerr black hole. It furnishes another encouraging tool to estimate black hole parameters and test theories of gravity in extreme regions near the event horizon. We propose a technique that uses EHT observables, the angular shadow diameter $d_{sh}$ and the axis ratio $\mathcal{D}_A$, to estimate the parameters associated with SMBHs, described by the Kerr metric. Unlike previous methods, our approach explicitly considers the uncertainties in the measurement of EHT observables. Modelling Kerr--Newman and three rotating regular spacetimes to be M87* and Sgr A* and applying our technique, we estimate the associated charge parameters along with spin. Our method is consistent with the existing formalisms and can be applied to shadow shapes that are more general and may not be circular. We can use the technique for other SMBHs once their EHT observables become accessible. With future, more accurate measurements of the EHT observables, the estimation of various SMBH parameters like the spin and inclination angles of M87* and Sgr A* would be more precise.
\end{abstract}

% Select between one and six entries from the list of approved keywords.
% Don't make up new ones.

%%%%%%%%%%%%%%%%%%%%%%%%%%%%%%%%%%%%%%%%%%%%%%%%%%

%%%%%%%%%%%%%%%%% BODY OF PAPER %%%%%%%%%%%%%%%%%%

\section{Introduction}\label{Section:Introduction}
Supermassive black holes (SMBHs), which are some of the most elusive objects in the Universe, provide a crucial link between the theoretical predictions and observational findings of several physical phenomena like the formation and co-evolution of the host galaxies \citep{1998Natur.395A..14R,2013ARA&A..51..511K,2019NatAs...3...48S}, tidal disruption of companion astronomical objects \citep{2017NatAs...1E..33L,Ryu:2022xkw} and star formation \citep{2022Natur.601..329S}. However, as the most vital sources of gravity, they may hold the key to solving more complex fundamental questions, such as finding a unified theory of gravity adept at explaining the Universe's workings at different length scales, at both low and high energies. According to the no-hair theorem \citep{Carter:1971zc} of general relativity (GR), three parameters, namely the black hole mass, spin and electric charge \citep{Israel:1967wq,Hawking:1971vc} uniquely describe astrophysical BHs. The deviations from the Kerr black hole metric, the Kerr-like metrices, may arise because of a plethora of modified theories of gravity (MTGs) 
\citep{2008arXiv0807.1640L,Nojiri:2017ncd,Capozziello:2011et,2022arXiv220507787V} -- each aimed at overcoming the anomalies arising off GR. They may also arise because of various surrounding matter-energy distributions \citep{Chen:2005qh,Atamurotov:2015nra, Ghosh:2014pga,Afrin:2021ggx,Anjum:2023axh}. These metrices may be endowed with additional deviation (charge or hair) parameter(s) which may characterize them in a theory agnostic framework \citep{Johannsen:2011dh,Konoplya:2016jvv}. Precise estimation of the black hole parameters is essential to describe the SMBHs and access fundamental signatures of the background theory of gravity. 

Multiple observational techniques can estimate the mass and distance of the BHs; the commonly used primary methods include stellar dynamical estimates \citep{2003ApJ...596.1015S,2021ApJ...916...25R}, modelling velocity dispersion of a Keplerian disk of gas, i.e., gas dynamical methods \citep{2013ApJ...770...86W,2015ApJ...809..101D} and the reverberation mapping of emission lines \citep{2010ApJ...721..715D,2020A&A...642A..59R}. The secondary methods include exploiting relationships between the
black hole mass and stellar bulge velocity dispersion \citep{2000ApJ...539L...9F,2009ApJ...698..198G} and host-galaxy bulge luminosity \citep{1998AJ....115.2285M}. However, there can be discrepancies between mass estimates obtained by the different methods, sometimes amounting to large orders of differences \citep{2013ApJ...770...86W}. While masses of SMBHs are relatively more straightforward to estimate because of their large-scale gravitational influence, the effects of the black hole spin are more difficult to measure because their general relativistic effects are significant only closest to the black hole and diminish outward \citep{2009ApJ...699..513L}. The most widely used techniques for estimating black hole spin are the analysis of the iron K$\alpha$ line \citep{2013mams.book.....B,2021MNRAS.504.3424F} and the thermal continuum-fitting method \citep{McClintock:2013vwa,2014SSRv..183..277R}, however, both approaches are astrophysical model dependent, have parametric degeneracy \citep{2020ApJ...897...84T} and are effective only for limited energy ranges, besides being prone to observational inaccuracies incurred by low signal-to-noise ratios \citep{2018A&A...614A..44K}. 

Among the numerous direct and indirect methods available for estimating black hole parameters, we can expect that no single approach can accurately measure all intrinsic and extrinsic parameters. It is anticipated that only through a combination of different independent measuring techniques can a comprehensive description of the SMBHs be achieved. Thus, developing novel and more robust strategies for the estimation of the black hole parameters is imperative, which, besides being independent of the astrophysical phenomena, should additionally be repeatable over multiple temporal epochs for the same source SMBH \citep{2021A&A...650A..56R}. Spatially resolving the region in and around $\sim 10\, G M/c^2$ off the black hole centre offers one such prospect \citep{2018A&A...614A..44K}, and is possible with radio observations \citep{2008Natur.455...78D}.

With the first horizon-scale images of the SMBHs M87* \citep{EventHorizonTelescope:2019dse,EventHorizonTelescope:2019ggy} and Sgr A* \citep{EventHorizonTelescope:2022xnr,EventHorizonTelescope:2022xqj} by the Event Horizon Telescope (EHT) collaboration, the black hole shadow -- a purely strong field construct independent of astrophysical phenomena -- has now opened new avenues to test various MTGs, and is also an up-and-coming tool to develop new techniques of black hole parameter estimation. Since the shadow silhouette is a strong field construct, the shape and size have been quantified into several observable quantities. They have been used for estimating the black hole parameters---Hioki \& Maeda (\citeyear{Hioki:2009na}) gave a numerical method to estimate the spin and inclination angle of Kerr black holes from the shadow radius $R_s$ and distortion $\delta_s$ which was extended analytically by Tsupko (\citeyear{Tsupko:2017rdo}), further, distortion features of the shadow was obtained analytically in a coordinate-independent manner \citep{Abdujabbarov:2015xqa} and later, Kumar \& Ghosh (\citeyear{Kumar:2018ple}) used three other shadow observables, namely the shadow area, circularity deviation and oblateness, that do not demand any specific shadow symmetry. Both the prescriptions have been widely used to estimate the spin and other charges of rotating black holes \citep{Kumar:2017tdw,Afrin:2021imp,Afrin:2021wlj,Afrin:2022ztr} and, recently Afrin \& Ghosh (\citeyear{Afrin:2021ggx}) have estimated the cosmological constant from the black hole shadow.
We aim to use two shadow observables obtained by the EHT for the SMBHs M87* and Sgr A*. We consider the possibility of  estimating various intrinsic black hole parameters like spin and charge alongside the extrinsic parameter, viz., inclination angle subject to the various observational uncertainties and assess the dependence of the estimated parameters on future more accurate observations, viz., with $\lesssim1\%$ error. We demonstrate the parameters obtained through our prescription align with previous estimates derived from independent methods. Furthermore, the technique can estimate the parameters of other SMBHs when their EHT observables are available.

This paper is structured as follows: Section~\ref{Section:2} establishes a theoretical framework for the estimation method by providing an analytical description of black hole shadows and introducing three well-motivated charged rotating regular black hole models. Section~\ref{Section:3} introduces the EHT observables and elucidates the parameter estimation technique. We demonstrate the method's applicability in estimating the charges of three rotating regular black hole models. Finally, Section~\ref{Sec:Conclusion} summarizes our findings and discusses future prospects.
%%%%%%%%%%%%%%%%%%%%%%%%%%%%%%%%%%%%%%%%%%%%%%%%%%%%%%
\section{Black hole shadow: Analytic description}\label{Section:2}
\begin{figure*}
\begin{center}
    \begin{tabular}{c c}
     \includegraphics[scale=0.90]{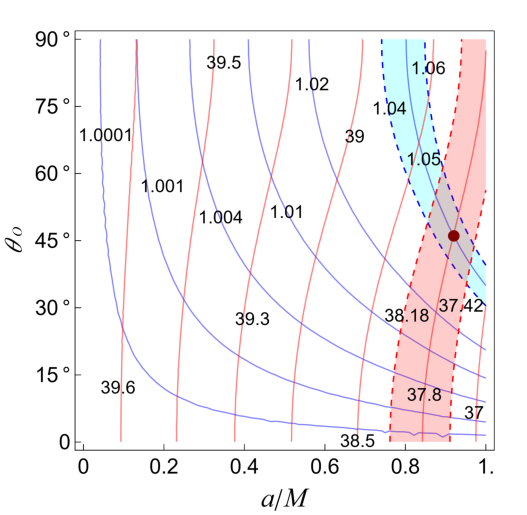}&
     \includegraphics[scale=0.90]{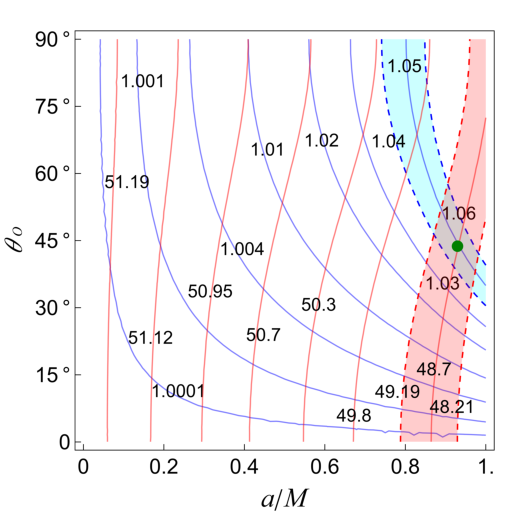}
\end{tabular}
\end{center}
	\caption{Contour plots of the EHT observables $d_{sh}$ and $\mathcal{D}_A$ in the plane $(a, \theta_0)$ of a Kerr black hole. We label each curve with the corresponding values of $d_{sh}$ (red lines) and $\mathcal{D}_A$ (blue lines). The shaded regions correspond to $\lesssim1\%$ error bars of EHT results. The brown point (left) and the green point (right) denote the estimated spin $a$ and the inclination $\theta_0$ of SMBHs M87* and Sgr A*, respectively.}
	\label{Fig:Estimation_Kerr}
\end{figure*}
%%%%%%%%%%%%%%%%%%%%%%%%%%%%%%%%%%%%%%%%%%%%%%%%%%%%%%%%%%%%%%%%
The black hole shadow is a dark brightness depression on the observer's sky, outlined by a bright ring-like structure \citep{1979A&A....75..228L,Johnson:2019ljv}. We begin by analysing the Kerr-like metric's shadow features -- an asymptotically flat, stationary, and axisymmetric spacetime whose line element in Boyer-Lindquist coordinates reads \citep{Bambi:2013ufa,Tsukamoto:2017fxq,Kumar:2020yem}
\begin{eqnarray}\label{metric}
ds^2 & = & - \left( 1- \frac{2m(r)r}{\Sigma} \right) dt^2  - \frac{4am(r)r}{\Sigma  } \sin^2 \theta\, dt \, d\phi +
\frac{\Sigma}{\Delta}dr^2  \nonumber
\\ && + \Sigma\, d \theta^2+ \left[r^2+ a^2 +
\frac{2m(r) r a^2 }{\Sigma} \sin^2 \theta
\right] \sin^2 \theta\, d\phi^2,
\end{eqnarray}
where
\begin{equation}
\Sigma = r^2 + a^2 \cos^2\theta,\;\;\;\;\;  \Delta(r)= r^2 + a^2 - 2m(r)r,\label{Delta}
\end{equation}
and $m(r)$ is the mass function, which can also be a be an explicit function of other black hole charges, such that $\lim_{r\to\infty}m(r)=M$; here $M$ is the ADM mass of the Kerr black hole and $a$ is the spin parameter.The Kerr-Newman black hole and three rotating regular black holes can be obtained by choosing the $m(r)$ appropriately.

It turns out that  $\Delta(r)=0$ has one or more positive roots with the largest root $r_+$ denoting the event horizon radius, and that $m(r)$ is well behaved for $r>r_+$.
Like the Kerr black hole, the black hole (\ref{metric}) possesses the Killing vectors $\chi_{(t)}^{\mu}=\delta _t^{\mu }$  and $\chi_{(\phi)}^{\mu}=\delta _{\phi }^{\mu }$ owing to the time translational and rotational invariance isometries respectively.
Due to the spacetime symmetries, the metric (\ref{metric}), admits conserved quantities, energy $\mathcal{E}=-p_t$ and axial angular momentum $\mathcal{L}_z=p_{\phi}$, where $p_{\mu}$ is the photon's four-momentum. Apart from these, the Carter's constant $\mathcal{K}$ \citep{Carter:1968rr,Chandrasekhar:1985kt} ensures that the null geodesic equations become completely separable in the Kerr-like spacetime \citep{Tsukamoto:2017fxq,Kumar:2018ple,Kumar:2020ltt}. We are interested in the radial geodesic equation which is given by 
\begin{align}
\Sigma \frac{dr}{d\lambda}&=\pm\sqrt{\mathcal{R}(r)}\ ,\label{RadialEq}\\
\mathcal{R}(r)&=\left((r^2+a^2){\mathcal{E}}-a{\mathcal {L}_z}\right)^2-\Delta({\cal K}+(a{\mathcal{E}}-{\mathcal {L}_z})^2)\label{rpot},
\end{align}
where $\mathcal{R}(r)$ is the radial potential.
The black hole shadow silhouette is formed by the locus of spherical photons orbits (SPOs) with constant radii $r_p^\mp$ obtained by solving $\mathcal{R}(r_p^\mp)=\mathcal{R}'(r_p^\mp)=0$, which yield the critical values of impact parameters $\xi\equiv \mathcal{L}/\mathcal{E},\,\, \eta\equiv \mathcal{K}/\mathcal{E}^2,$ \citep{Tsukamoto:2017fxq,Kumar:2018ple,Kumar:2020yem} as
\begin{align}
\xi_c=&\frac{[a^2 - 3 r_p^2] m(r_p) + r_p [a^2 + r_p^2] [1 + m'(r_p)]}{a [m(r_p) + r_p [-1 + m'(r_p)]]},\nonumber\\
\eta_c=&-\frac{r_p^3}{a^2 [m(r_p) + r_p [-1 + m'(r_p)]]^2}\Big[r_p^3  \nonumber\\
&+ 9 r_p m(r_p)^2+ 2 [2 a^2 + r_p^2+r_p^2 m'(r_p)] \nonumber\\
& \times r_pm'(r_p)-  2 m(r_p) [2 a^2 + 3 r_p^2 + 3 r_p^2 m'(r_p)]\Big].\label{impactparameter}
\end{align}
Equation~(\ref{impactparameter}) recovers the critical impact parameters around the Kerr black hole, i.e.,  when $m(r)=M$ \citep{Kumar:2018ple, Afrin:2021imp}.
We can trace the gravitationally lensed image of the photon region on the celestial sky of an asymptotically faraway observer ($r_0\to\infty$), making an inclination angle $\theta_0$ with the black hole axis that would outline a dark interior region, namely the black hole shadow \citep{Johannsen:2015mdd,Johnson:2019ljv}, whose coordinates are obtained as \citep{Bardeen:1973tla,Kumar:2020yem,Afrin:2021wlj}
\begin{equation}
\{X,Y\}=\{-\xi_{c}\csc\theta_0,\, \pm\sqrt{\eta_{c}+a^2\cos^2\theta_0-\xi_{c}^2\cot^2\theta_0}\}\,.\label{pt}
\end{equation}

The shadow boundary is purely a strong field construct, entirely determined by the spacetime geometry, which makes it an ideal tool to extract information about the background theory of gravity \citep{Johannsen:2010ru,Baker:2014zba,Cunha:2018acu}, or more precisely, the intrinsic parameters of the target SMBH; but, the extrinsic parameters, e.g., the inclination angle $\theta_0$ can also be extracted. Thus, the characteristic properties of the shadow, if astrophysically resolvable, can give a theoretically clean parameter estimation method that is expected to be unaffected or only sub-dominantly affected by any additional astrophysical events, including the plasma and accretion dynamics \citep{Johannsen:2015mdd,Johnson:2019ljv,Ricarte:2022kft}. However, we must remember that this would be observationally hard to achieve compared to other astrophysical methods involving the plasma structure and accretion dynamics \citep{Ricarte:2022kft}. Further, like the mass and spin parameters, other intrinsic parameters, e.g., charge parameters, also influence the shadow characteristics in a measurable amount and thus can be retrieved from the shadow structure. 
%%%%%%%%%%%%%%%%%%%%%%%%%%%%%%%%%%%%%%%%%%%%%%%%%%%%%%%%%%
\begin{figure*}
\begin{center}
    \begin{tabular}{c c}
     \includegraphics[scale=0.90]{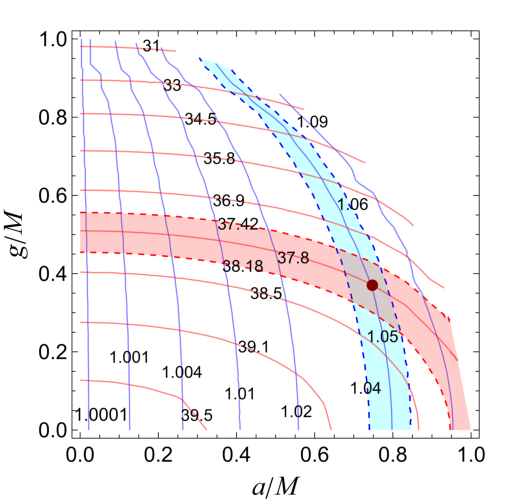}&
     \includegraphics[scale=0.90]{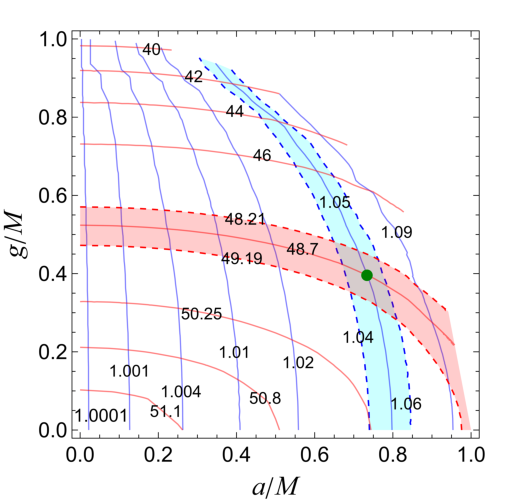}
\end{tabular}
\end{center}
	\caption{Contour plots of the EHT observables $d_{sh}$ and $\mathcal{D}_A$ in the plane $(a, g)$ for a Kerr--Newman black hole. We label each curve with the corresponding values of $d_{sh}$ (red lines) and $\mathcal{D}_A$ (blue lines). The brown point (left) and the green point (right) denote the estimated spin $a$ and the charge $g$ of SMBHs M87* and Sgr A*, respectively.}
	\label{Fig:Estimation_KN}
\end{figure*}
%%%%%%%%%%%%%%%%%%%%%%%%%%%%%%%%%%%%%%%%%%%%%%%%%%%%%%%%%%

We will explore the above possibility with various rotating charged black holes focusing on four well-motivated models sourced by electric and non-linear electrodynamics (NED) charges, viz., electrically charged Kerr--Newman black holes and magnetically charged rotating Bardeen, Hayward and Ghosh-Culetu black holes. We shall assess the possibility of estimating the charge and spin parameters of M87* and Sgr A*.

\paragraph{Kerr--Newman black holes}
{Despite the long-standing belief of charge neutrality \citep{Gibbons:1975kk}, there are multiple studies suggesting that astrophysical black holes may have charge induced \citep{Wald:1974np,2004astro.ph..5237D,Levin:2018mzg} or it may get accumulated \citep{2019Obs...139..231Z}. Furthermore, the charge neutrality of SMBHs is yet to be observationally established, and recently, there have been a lot of efforts to constrain the electric charge of SMBHs. The mass function of the Kerr--Newman black hole is given by
\begin{equation}\label{KN_mass}
    m(r)= M-\frac{g^2}{2r},
\end{equation}
where $g$ is the electric charge, we retrieve the Kerr black hole for $g=0$.
Recently, the EHT collaboration has obtained the constraint $g\in(0,0.90M]$ for M87* \citep{EventHorizonTelescope:2021dqv}; the charge of Sgr A* has been theoretically limited to $g\leq 3.1 \times 10^8$C using Chandra X-ray observations \citep{Karouzos2018,Zajacek:2018ycb} and $g\in (0, 0.7174M]$ using the EHT results for Sgr A* \citep{Ghosh:2022kit}. Besides other recent studies, these highlight the relevance of determining the electric charge.}

\paragraph{Rotating Bardeen black holes}
{Next, we shall consider some regular rotating black hole models that address the central singularity problem in classical GR black holes \citep{1966JETP...22..241S}. Here, we shall implicitly assume the accreting matter to be non-interacting with the background matter around the black holes.
Bardeen (\citeyear{Bardeen:1968}) proposed the first regular
asymptotically ($r\to\infty$) flat black hole with a de-Sitter core
having an equation of state $P=-\rho$, having horizons and no curvature singularity \citep{1966JETP...22..241S}. It is an exact solution of the Einstein field equations coupled with nonlinear electrodynamics (NED) associated with the magnetic monopole charge $g$ \citep{Ayon-Beato:2000mjt}. The rotating Bardeen black holes can be obtained from metric (\ref{metric}) with the mass function $m(r)$ given by \citep{Bambi:2013ufa,Kumar:2020yem}
\begin{equation}
m(r)=M\left(\frac{r^2}{r^2 + g^2}\right)^{3/2}.
\end{equation}
The Kerr black hole is recovered without the NED ($g=0$). With increasing charge parameter $g$, the shadows of rotating Bardeen black holes become smaller and more distorted \citep{Abdujabbarov:2016hnw,Kumar:2020ltt,KumarWalia:2022aop}.
Using X-ray data from the disk surrounding
the black hole candidate in Cygnus X-1, the rotating Bardeen black hole metric has been put to the test, such that the bounds on black hole parameters $a>0.78M$ and $g<0.41M$,
and $a>0.89M$ and $g<0.28M$ \citep{Bambi:2014nta} are obtained. The shadows of Bardeen black holes have been found entirely consistent with the EHT results for  M87* \citep{EventHorizonTelescope:2021dqv} and, for $a = 0$, $0.18M\leq g\leq0.595M$ and $a=0.2M$, $0.154M\leq g\leq0.58367M$, they are consistent with the EHT results for the SMBH Sgr A* \citep{KumarWalia:2022aop}.}
%%%%%%%%%%%%%%%%%%%%%%%%%%%%%%%%%%%%%%%%%%%%%%%%%%%
\begin{figure*}
\begin{center}
    \begin{tabular}{c c}
     \includegraphics[scale=0.90]{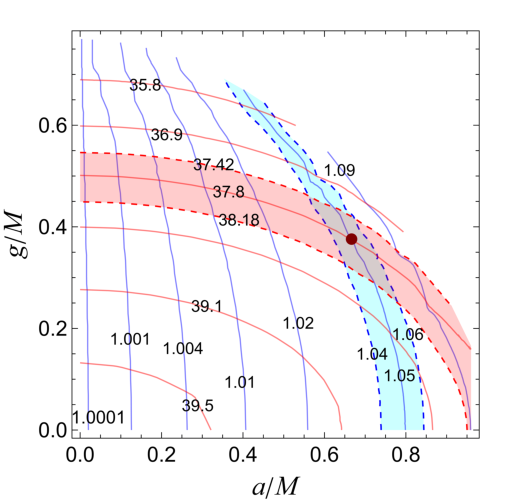}&
     \includegraphics[scale=0.90]{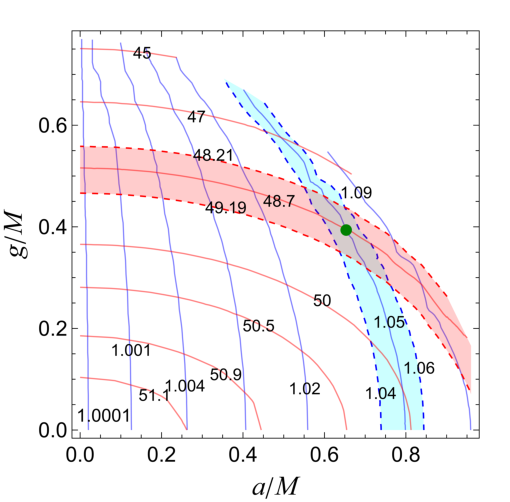}
\end{tabular}
\end{center}
	\caption{Contour plots of the EHT observables $d_{sh}$ and $\mathcal{D}_A$ in the plane $(a, g)$ of a rotating Bardeen black hole. We label each curve with the corresponding values of $d_{sh}$ (red lines) and $\mathcal{D}_A$ (blue lines). The brown point (left) and the green point (right) denote the estimated spin $a$ and the charge $g$ of SMBHs M87* and Sgr A*, respectively.}
	\label{Fig:Estimation_Bardeen}
\end{figure*}
%%%%%%%%%%%%%%%%%%%%%%%%%%%%%%%%%%%%%%%%%%%%%%%%%%%%%%%%%%%%

\paragraph{Rotating Hayward black holes}
{Hayward (\citeyear{Hayward:2005gi}) obtained another important and well-studied regular black hole solution. Besides the mass $M$, the Hayward black hole has one
additional parameter $g$  related to the magnetic monopole charge  \citep{Fan:2016hvf}. The mass function described the rotating Hayward black hole \citep{Bambi:2013ufa,Kumar:2020yem}
\begin{equation}
m(r)=\frac{Mr^3}{r^3+g^3}.\label{Eq:Hayward_MassFunc}
\end{equation}
A systematic bias analysis has revealed that the shadows of rotating Hayward black holes degenerate with those cast by the Kerr black holes for small values of the charge parameter $g$ \citep{Kumar:2020yem}. From the shadow analysis of rotating Hayward black holes and imposing the EHT bounds on the shadow asymmetry $\leq0.1$ of M87*, the upper limit  $g\leq 1.0582M$ could be placed whereas, the angular diameter of M87* constrains the charge to be $g\leq 0.73627M$ \citep{Kumar:2020yem}. The EHT results for Sgr A* puts limit $g\geq0.5302M$ on the charge \citep{KumarWalia:2022aop}.} 

\paragraph{Ghosh-Culetu black holes}
{Unlike Bardeen and Hayward black holes, which have asymptotically de-Sitter cores, the Ghosh-Culetu black holes \citep{Ghosh:2014pba}, a novel class of rotating regular black holes have asymptotic Minkowski core \citep{Simpson:2019mud} and are described by metric \eqref{metric} with mass function \citep{KumarWalia:2022aop}
\begin{equation}
m(r)=Me^{-g^2/2Mr}.\label{Eq:Ghosh_MassFunc}
\end{equation}  
The Ghosh-Culetu black hole was obtained by Ghosh \citep{Ghosh:2014pba} by generalizing the spherically symmetric counterpart \citep{Culetu:2014lca}. Whilst these black holes asymptotically go over to the Kerr-Newman black holes, the Kerr black hole is recovered for $g=0$. Though the Ghosh--Culetu black holes possess several properties common to the rotating Bardeen and Hayward black holes, there are significant deviations mainly in the deep core region \citep{Simpson:2019mud}. The size of the shadow cast decreases, and the distortion in shadow shape increases
with increasing charge $g$ \citep{KumarWalia:2022aop}. 
The shadow asymmetry of M87* barely constrains the charge parameter to be $g\leq1.2130M$ and $d_{sh}^{M87^*}$ puts the upper bar $g\leq0.30461M$ \citep{Kumar:2020yem}. The EHT results for Sgr A* put constraints $0.18345M\leq g\leq
0.62058M$ at $a = 0$, and $0.155M\leq g\leq 0.61116M$ at $a = 0.20M$ \citep{KumarWalia:2022aop}.}

\section{Parameter estimation via EHT observables}\label{Section:3}
The shadow characteristics, i.e., the size and, more importantly, the shape, are distinguishing observable features of the SMBHs, and are dominantly affected by the parameters of the background metric; quantifying these observables is our first step towards estimating the SMBH parameters.
The effect of spin on the shadow shape is marginal for pole-on inclinations \citep{Ricarte:2022kft}, while for higher inclinations, both the shadow size and shape change as the shadow grows smaller by up to 7.5\% \citep{EventHorizonTelescope:2022xqj}, and is more distorted. Similar effects on the shadow characteristics are caused by additional hairs \citep{Cunha:2015yba,Afrin:2021imp,Zhang:2022osx,Anjum:2023axh} for a given $a$ and $\theta_0$---this prompts the use of suitable shadow observables for estimating both the intrinsic and extrinsic black hole parameters \citep{Hioki:2009na,2013ApJ...777...13C,Wei:2019pjf,Kumar:2018ple,Afrin:2021imp,Afrin:2021ggx}.
Proper characterization of the shadow shape and size is imperative and several shadow observables have been proposed---Hioki \& Maeda (\citeyear{Hioki:2009na}) numerically estimated $a$ and $\theta_0$ from the shadow radius $R_s$ and the distortion $\delta_s$, which was extended analytically by Tsupko (\citeyear{Tsupko:2017rdo}). However, $\delta_s$ is degenerate in $a$ and deviations parameters from 
Kerr \citep{Kumar:2018ple}, besides, both observables demand some specific symmetry in the shadow shape, and hence might not be adequately effective in some MTGs \citep{Abdujabbarov:2015xqa,Tsukamoto:2014tja,Kumar:2018ple}. To overcome this shortcoming Kumar \& Ghosh (\citeyear{Kumar:2018ple}) utilized two other observables, namely the shadow area given by
\begin{eqnarray}
A&=&2\int_{r_p^{-}}^{r_p^+}\left( Y(r_p) \frac{dX(r_p)}{dr_p}\right)dr_p,\label{Area}
\end{eqnarray} 
%%%%%%%%%%%%%%%%%%%%%%%%%%%%%%%%%%%%%%%%%%%%%%%%%%%%%%%
\begin{figure*}
\begin{center}
    \begin{tabular}{c c}
     \includegraphics[scale=0.90]{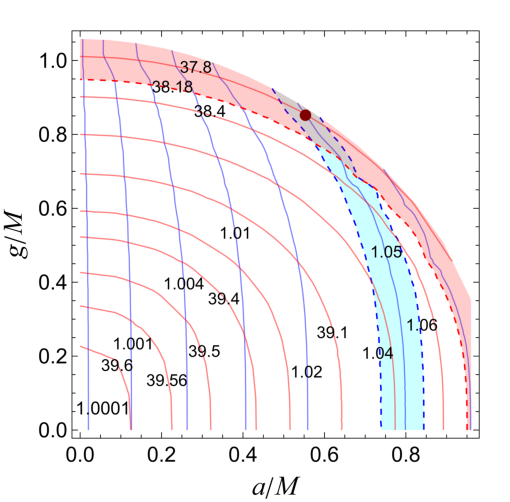}&
     \includegraphics[scale=0.90]{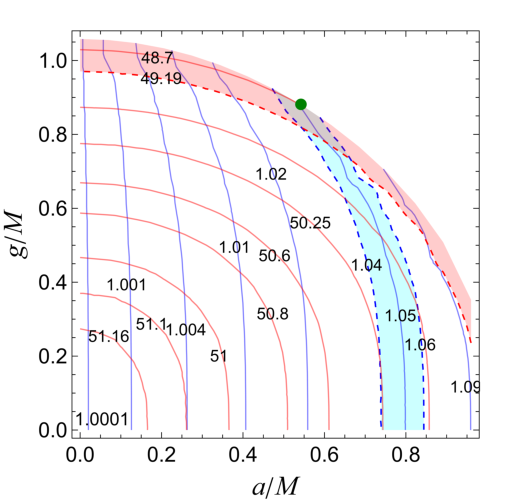}
\end{tabular}
\end{center}
	\caption{Contour plots of the EHT observables $d_{sh}$ and $\mathcal{D}_A$ in the plane $(a, g)$ of a rotating Hayward black hole. We label each curve with the corresponding values of $d_{sh}$ (red lines) and $\mathcal{D}_A$ (blue lines). The brown point (left) and the green point (right) denote the estimated spin $a$ and the charge $g$ of SMBHs M87* and Sgr A*, respectively.}
	\label{Fig:Estimation_Hayward}
\end{figure*}
%%%%%%%%%%%%%%%%%%%%%%%%%%%%%%%%%%%%%%%%%%%%%%%%%%%%%%%%%%
and the ratio of the horizontal to the vertical shadow diameters $D$ for estimating the black hole parameters from any generic haphazard shadow shape. Subsequently, the Kumar \& Ghosh method led to estimating the parameters of several rotating BHs in MTGs with this prescription \citep{Kumar:2017tdw,Afrin:2021imp,Afrin:2021wlj,Afrin:2021ggx,Afrin:2022ztr}. However, these estimation methods require precise measurements of the shadow observables. Additionally a precise measurement of the black hole's mass, distance, etc. is required; also, the various measurement/observational uncertainties have not been adequately quantified. Here, we aim to tackle the issue and consider the errors in measuring the EHT observables in our analysis.

We now outline our estimation technique. First, we introduce two EHT observables: the angular diameter of the shadow $d_{sh}$ and the axis ratio $\mathcal{D}_A$. While $d_{sh}$ quantifies the size of the shadow, the deformation in the shadow shape is quantified by $\mathcal{D}_A$; hence, the two observables together give a nearly complete description of the black hole shadow. Supposing the target black hole to be at a distance $d_{bh}$ from Earth, the angular diameter of the shadow cast is given by \citep{Ghosh:2022kit}
\begin{eqnarray}
d_{sh}=2\frac{R_a}{d_{bh}}\;,\;R_a=\sqrt{A/\pi},\label{angularDiameterEq}
\end{eqnarray}  
where $bh\equiv\{$M87*, Sgr A*$\}$, respectively, for the two SMBHs, however, $bh$ can be any other target SMBH. The images of both M87* and Sgr A* exhibit a thick luminous emission ring, with inferred diameters (42 $\pm$ 3)$\mu$as and (51.8 $\pm$ 2.3)$\mu$as, respectively \citep{EventHorizonTelescope:2019dse,EventHorizonTelescope:2022xnr} surrounding a region of brightness depression, namely the black hole shadow \citep{EventHorizonTelescope:2019dse,EventHorizonTelescope:2022xqj}. For M87*, there is a $\lesssim10\%$ offset between the measured emission ring diameter and the size of the photon ring \citep{EventHorizonTelescope:2019ggy}, which thus leads to the calibrated value of the angular diameter $d_{sh}^{M87^*}=37.8\mu$as, considering a $10\%$ offset from the mean ring diameter. For Sgr A*, however, the EHT collaboration has  calibrated the emission ring diameter to the shadow diameter and obtained the value of angular diameter (48.7$\pm$7)$\mu$as \citep{EventHorizonTelescope:2022xnr,EventHorizonTelescope:2022xqj}. We shall consider the mean shadow diameter $d_{sh}^{Sgr A^*}=48.7\mu$as for our estimation.
We note that $d_{sh}$ implicitly depends on the black hole mass beside the distance of the SMBH, as the observable $A$ has a dimension of $M^2$. 

While the $d_{sh}$ correlates the mass and distance of the SMBH to shadow size, the deviation from a circularity may directly hint at both a nonzero inclination angle and a nonzero black hole spin. Next, we see that the circular asymmetry in the shadows cast by M87* and Sgr A* can be gauged in terms of the axial ratio $\mathcal{D}_A$, i.e., the ratio of major to minor diameters of the shadow \citep{EventHorizonTelescope:2019dse} which is given by \citep{Afrin:2021imp}
\begin{equation}
    \mathcal{D}_A=\frac{\Delta Y}{\Delta X},
\end{equation}
The EHT has measured the fractional spread in the emission ring radius from the reconstructed images of M87* and found it highly symmetric with the axis ratio constrained to be $1<\mathcal{D}_A\lesssim 1.33$  \citep{EventHorizonTelescope:2019dse,EventHorizonTelescope:2019ggy}. Recently, Tiede et al. (\citeyear{Tiede:2022bdd}) have obtained the ellipticity $\tau$ of the shadow of M87* to be $\tau\in[0, 0.5]$ amounting to an axis ratio $\sim 2:1$. However, owing to the sparse interferometric coverage and significant uncertainties in the circularity measurement, no results for axis ratio have been released for the shadow of Sgr A* \citep{EventHorizonTelescope:2022xqj}. Still, a precise measurement of $\mathcal{D}_A$ with an expanded array of telescopes is possible. The Kerr black holes permit only $1\leq\mathcal{D}_A\leq1.1$ \citep{Tsupko:2017rdo}, and we shall consider the mean value $\mathcal{D}_A=1.05$ for both M87* and Sgr A* to demonstrate our method.

Having set up the EHT observables, we can estimate the parameters of SMBHs using a simple contour intersection \citep{Kumar:2018ple}. However, as discussed earlier, there are several caveats to our analytical prescription of parameter estimation because of various underlying uncertainties in the observations, which also affect the measurements of mass and distance of the SMBHs. But, to derive the bounds on observables, the EHT results already consider many of these uncertainties \citep{Afrin:2021wlj}. We would thus consider the inferred priors on the masses of the M87* and Sgr A* to be $M_{M87^*}= 6.5\times 10^9 M_\odot$ and $M_{SgrA^*} = 4.0 \times 10^6 M_\odot $ \citep{EventHorizonTelescope:2019dse,EventHorizonTelescope:2019pgp,EventHorizonTelescope:2022xnr,EventHorizonTelescope:2022xqj} respectively, and their respective distances from Earth to be $d_{M87^*}=16.8$Mpc and $d_{SgrA^*} = 8$kpc \citep{EventHorizonTelescope:2019dse,EventHorizonTelescope:2019pgp,EventHorizonTelescope:2022xnr,EventHorizonTelescope:2022xqj}. 

But, the current measurement errors in the directly measured values of the EHT observables are quite large, viz., $\sim14\%$ error in the $d_{sh}$ for Sgr A*. It is difficult to place any tight constraints, let alone estimate the SMBH parameters at the present resolution of the EHT. If we could measure shadow diameter and asymmetry with $\sim1\%$ accuracy, the estimation of inclination and spin would be feasible. Besides, since the EHT measures ratio of the shadow diameter to the source distance, hence a $\sim1\%$ measurement of the shadow diameter would require an independent $\sim1\%$ accurate distance measurement--which is very challenging, particularly for M87*. However, with future, more precise observational capabilities of the  next-generation EHT (ngEHT) and space VLBI \citep{Ricarte:2022kft,2023Galax..11....2B}, and also independent measurements of the black holes mass-to-distance ratio, viz., by the stellar dynamical observations with Sgr A*, obtaining estimations of various SMBH parameters via EHT observables can become possible. To explore this possibility, we shall consider a $\lesssim 1\%$ error bar on the EHT observable values---which may be accomplished with future observations---to carry out our estimation and show the dependence of the estimated results on the measurement errors.
%%%%%%%%%%%%%%%%%%%%%%%%%%%%%%%%%%%%%%%%%%%%%%%%%%%%%%%%%
\begin{figure*}
\begin{center}
    \begin{tabular}{c c}
     \includegraphics[scale=0.90]{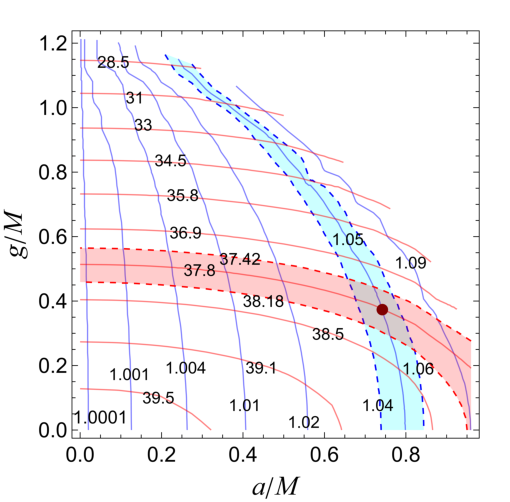}&
     \includegraphics[scale=0.90]{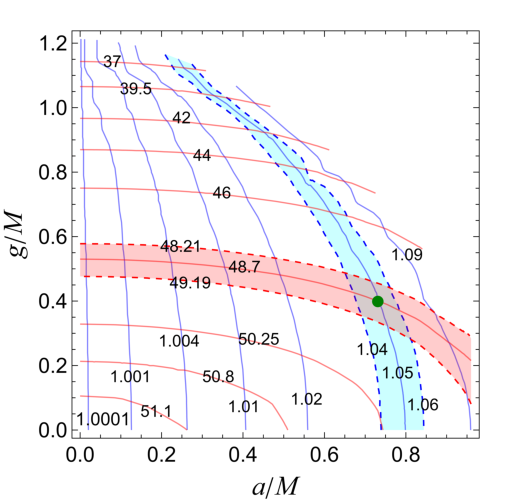}
\end{tabular}
\end{center}
	\caption{Contour plots of the EHT observables $d_{sh}$ and $\mathcal{D}_A$ in the plane $(a, g)$ of a Ghosh--Culetu black hole. We label each curve with the corresponding values of $d_{sh}$ (red lines) and $\mathcal{D}_A$ (blue lines). The brown point (left) and the green point (right) denote the estimated spin $a$ and the charge $g$ of SMBHs M87* and Sgr A*, respectively.}
	\label{Fig:Estimation_Ghosh}
\end{figure*}
%%%%%%%%%%%%%%%%%%%%%%%%%%%%%%%%%%%%%%%%%%%%%%%%%%%%%%%%%%

%%%%%%%%%%%%%%%%%%%%%%%%%%%%%%%%%%%%%%%%%%%%%%%%%%%%%%%%%%%%%%%%%%%%%
\begin{table}
\caption{Estimated parameters of SMBHs modelled as Kerr black holes using EHT observables.}
      \centering
{\renewcommand{\arraystretch}{1.4}      
\begin{tabular}{c|cc}
\hline\hline
\multirow{2}{*}{SMBH} & \multicolumn{2}{c}{Estimated parameters}                      \\ \cline{2-3} 
                      & \multicolumn{1}{c|}{$\theta_0$}                       & $a/M$  \\ \hline
M87*                  & \multicolumn{1}{c|}{$46^{\circ+15.3^\circ}_{-10.3^\circ}$} & $0.91^{+0.05}_{-0.06}$ \\ \hline
Sgr A*                & \multicolumn{1}{c|}{$44^{\circ+14.1^\circ}_{-9.8^\circ}$} & $0.93^{+0.05}_{-0.06}$   \\ \hline\hline
\end{tabular}
}
\label{Table:Parmameters_Kerr}
\end{table}
%%%%%%%%%%%%%%%%%%%%%%%%%%%%%%%%%%%%%%%%%%%%%%%%%%%%%%%%%%%%%%%%%%%%%%%%
%\subsection{}
As the EHT results are consistent with general relativity (GR), we can model any SMBHs, such as M87* and Sgr A*, with the Kerr metric to estimate their parameters. 
As discussed earlier, there are independent estimates of the mass  of the SMBHs from several observations. If the distance to the SMBH is also known independently, then we only need to estimate one intrinsic parameter, the spin. We can also estimate the inclination angle since two shadow observables can estimate two black hole parameters. Here, we outline the technique to estimate these parameters.
\begin{itemize}
    \item Because of the effect incurred by $a$ and $\theta_0$ on the angular diameter of the shadow and the axis ratio, there may arise a possible degeneracy of the $d_{sh}$ and $\mathcal{D}_A$ in $a$ and $\theta_0$, such that, more than one combination of ($a$, $\theta_0$) may give the same $d_{sh}$ and/or $\mathcal{D}_A$.
    \item We examine the possibility of degeneracy in Fig.~\ref{Fig:Estimation_Kerr} by plotting constant contours of the shadow observables in the two-parameter subspace and find that a contour of an observable gives a one-to-one correspondence between the black hole parameters. In contrast, the contours of the two different observables intersect at unique points.
    \item We can readily infer that the shadow observables are degenerate for an infinite number of unique parameter points ($a, \theta_0$) lying on a constant contour curve. However, they are non-degenerate in the parameters ($a, \theta_0$) if at least one of the two parameters is fixed.
    \item From each intersection point of the contours of $d_{sh}$ and $\mathcal{D}_A$, one can uniquely determine the parameters $a$ and $\theta_0$ of the SMBHs.
    \item Thus, taking the intersection point of the $\mathcal{D}_A=1.01$ contour with the $d_{sh}^{M87^*}=37.8\mu$as contour (see the left panel in Fig.~\ref{Fig:Estimation_Kerr}) and/or  the $d_{sh}^{Sgr A^*}=48.7\mu$as contour (see the right panel in Fig.~\ref{Fig:Estimation_Kerr}) we estimate the spin and inclination angle of M87* and Sgr A* respectively. Further, from the intersection of the contours corresponding to the $1\%$ error in measured values of $d_{sh}$ and $\mathcal{D}_A$ (the dashed lines in Fig.~\ref{Fig:Estimation_Kerr}), we determine the uncertainties in the inferred parameters. We tabulate the estimated parameters in Table \ref{Table:Parmameters_Kerr}.
\end{itemize}  
The estimated spin $a=0.91^{+0.05}_{-0.06}M$ of M87* is consistent with earlier estimates, $a=(0.9\pm0.5)M$ from the measured photon angular momentum near M87* \citep{Tamburini:2019vrf}; however, the estimated inclination $\theta_0=46^{\circ+15.3^\circ}_{-10.3^\circ}$ shows a significant deviation from the estimated value $\theta_0\approx17$\textdegree ~by the EHT \citep{EventHorizonTelescope:2019dse,EventHorizonTelescope:2019ggy}. We estimate the spin and inclination angle of Sgr A* to be $a=0.93^{+0.05}_{-0.06}M$, $\theta_0=44^{\circ+14.1^\circ}_{-9.8^\circ}$, respectively. For Sgr A*, the EHT has disfavoured $\theta_0>50$\textdegree. Additionally, previous techniques have provided inconclusive estimates, ranging from low \citep{2009ApJ...706..960H,2011ApJ...735..110B,2016ApJ...820..137B,2020ApJ...901L..32F} to high  \citep{2009ApJ...706..497M,2012ApJ...755..133S} values, in their attempts to constrain the spin of Sgr A*. However, our estimation aligns with the higher end of the spin range, supporting the latter findings. These show the consistency of our results with those obtained by other independent probes. However, the observational systematics arising, e.g., because of observations made by different telescopes in the sparse array, relatively uncertain radiative and accretion phenomena \citep{Gralla:2020pra} etc., introduce uncertainties in the measured observables, which is expected to affect the estimated parameters as well. Our method infers SMBH parameters presuming a $\sim1\%$ accurate measurement of the EHT observables, and the inference accuracy would directly scale with the observational accuracy. Also, we can obtain the contour intersection point more accurately (greater number of significant digits) than the measured quantities. Still, we quote up to 3 significant digits in our results, considering the accuracy of the measured observables. Our estimation gives a good description of the SMBHs concurrent with the observational uncertainties of EHT. With future, more precise measurements, like with the ngEHT \citep{2023Galax..11....2B}, the method would yield an improved estimation of the SMBH parameters.
%%%%%%%%%%%%%%%%%%%%%%%%%%%%%%%%%%%%%%%%%%%%%%%%%%%%%%%%%%%%%%%%%%%
\begin{table*}
\caption{Estimated parameters of SMBHs using EHT observables.}
      \centering
{\renewcommand{\arraystretch}{1.3}      
\begin{tabular}{c|cccc}
\hline\hline
\multirow{2}{*}{Model} & \multicolumn{4}{c}{SMBH}                                                                       \\ \cline{2-5} 
                       & \multicolumn{2}{c|}{M87*}                                 & \multicolumn{2}{c}{Sgr A*}          \\ \cline{2-5} 
                       & \multicolumn{1}{c|}{$g/M$}  & \multicolumn{1}{c|}{$a/M$}  & \multicolumn{1}{c|}{$g/M$}  & $a/M$  \\ \hline
Kerr--Newman  & \multicolumn{1}{c|}{$0.37^{+0.09}_{-0.12}$} & \multicolumn{1}{c|}{$0.74^{+0.07}_{-0.08}$} & \multicolumn{1}{c|}{$0.40^{+0.08}_{-0.11}$}  & {$0.73^{+0.08}_{-0.08}$} \\ \hline
Rotating Bardeen                & \multicolumn{1}{c|}{$0.37^{+0.08}_{-0.14}$} & \multicolumn{1}{c|}{$0.66^{+0.11}_{-0.09}$} & \multicolumn{1}{c|}{$0.40^{+0.08}_{-0.11}$} & {$0.65^{+0.10}_{-0.10}$} \\ \hline
Rotating Hayward   & \multicolumn{1}{c|}{$0.85^{+0.06}_{-0.16}$} & \multicolumn{1}{c|}{$0.54^{+0.13}_{-0.07}$} & \multicolumn{1}{c|}{$0.89^{+0.03}_{-0.15}$} & {$0.53^{+0.13}_{-0.05}$} \\ \hline
Ghosh-Culetu & \multicolumn{1}{c|}{$0.37^{+0.09}_{-0.12}$} & \multicolumn{1}{c|}{$0.74^{+0.08}_{-0.08}$} & \multicolumn{1}{c|}{$0.40^{+0.08}_{-0.11}$} & {$0.73^{+0.07}_{-0.08}$} \\ \hline\hline
\end{tabular}
}
\label{Table:Parmameters_regular}
\end{table*}
%%%%%%%%%%%%%%%%%%%%%%%%%%%%%%%%%%%%%%%%%%%%%%%%%%%%%%%%%%%%%%%%%%%%%%%
Conversely, we can estimate the shadow observables if precise values of the SMBH parameters are known, e.g., considering $\theta_0=17$\textdegree and $a=0.9M$ for M87*, we estimate $d_{sh}=37.55\mu$as and $\mathcal{D}_A=1.0085$, and taking $\theta_0=50$\textdegree and $a=0.9M$ for Sgr A* gives $d_{sh}=49.08\mu$as and $\mathcal{D}_A=1.05$.
\subsection{Estimation of charge}
We gave a simple technique to estimate any two parameters of SMBHs. 
We have recently seen that the SMBHs M87* and Sgr A* can have charge and put upper limits on the charge parameter \citep{EventHorizonTelescope:2021dqv,EventHorizonTelescope:2022xqj,KumarWalia:2022aop,Ghosh:2022kit}. Here, we shall assume the SMBHs as rotating charged black holes to estimate the electric and magnetic charges. The EHT has other observational targets besides M87* and Sgr A*, and our method is applicable to all target SMBHs as and when results are available. We fix $\theta_0=90^\circ$ ~here for demonstrating the technique, as we can estimate only two parameters at a time, and independent acquirement of the inclinations of SMBHs would be required to estimate the charge and spin parameters.

Modeling the SMBHs as Kerr--Newman black holes and by using the contour intersection method in the ($a$, $g$) parameter space (see Fig.~\ref{Fig:Estimation_KN}), we obtain the estimates, $a=0.74^{+0.07}_{-0.08}M$ and $g=0.37^{+0.09}_{-0.12}M$ for M87* and, $a=0.73^{+0.08}_{-0.08}M$ and $g=0.40^{+0.08}_{-0.11}M$ for Sgr A*. Manifestly, both estimations are consistent within the earlier constrained ranges of spin parameters for the two SMBHs. For the rotating Bardeen black holes, we obtain the estimated parameters, $a=0.66^{+0.11}_{-0.09}M$, $g=0.37^{+0.08}_{-0.14}M$ for M87* and $a=0.65^{+0.10}_{-0.10}M$, $g=0.40^{+0.08}_{-0.11}M$ for Sgr A* respectively (see Fig.~\ref{Fig:Estimation_Bardeen}); clearly, these values are well within the earlier bounds on the parameters of the two SMBHs. For the rotating Hayward black hole, our estimation method yields the parameters of the SMBH M87* to be $a=0.54^{+0.13}_{-0.07}M$, $g=
0.85^{+0.06}_{-0.16}M$ and the parameters of Sgr A* to be $a=0.53^{+0.13}_{-0.05}M$, $g=0.89^{+0.03}_{-0.15}M$ (see Fig.~\ref{Fig:Estimation_Hayward}), which agree with the earlier bounds on the SMBH parameters.
With the Ghosh-Culetu black hole model, we obtain the estimated parameters of the SMBH M87* to be $a=0.74^{+0.08}_{-0.08}M$, $g=0.37^{+0.09}_{-0.12}M$ and the parameters of Sgr A* are estimated to be $a=0.73^{+0.07}_{-0.08}M$, $g=0.40^{+0.08}_{-0.11}M$ (see Fig. \ref{Fig:Estimation_Ghosh}) in agreement with prior constraints \citep{KumarWalia:2022aop}. We tabulate the inferred parameters in Table~\ref{Table:Parmameters_regular}. Although the estimated parameters obtained from the Kerr-Newman, Bardeen, and Ghosh-Culetu models yield consistent estimates for $a$ and $g$, the estimations derived from the Hayward black holes exhibit noticeable disagreement. Therefore, we adopt the values obtained from the former models, resulting in $g\approx0.37M$ for M87* and $g\approx0.40M$ for Sgr A*.
\section{Conclusion}\label{Sec:Conclusion}
The results of the Event Horizon Telescope (EHT) observations of the supermassive black holes (SMBHs) M87* and Sgr A* offer a new and compelling way to test various theories of gravity and to extract information about the black hole parameters, both extrinsic and intrinsic. Thus we can devise techniques for estimating the parameters of SMBHs.

We summarize the main results of our analysis below:
\begin{enumerate}
    \item Modeling the SMBHs M87* and Sgr A* as Kerr black holes, we estimate the spin and inclination of M87* to be $a=0.91^{+0.05}_{-0.06}M$ and $46^{\circ+15.3^\circ}_{-10.3^\circ}$ ~respectively. For Sgr A*, our estimation yields the corresponding values $a=0.93^{+0.05}_{-0.06}M$ and $\theta_0=44^{\circ+14.1^\circ}_{-9.8^\circ}$, respectively. 
     \item The estimated spin of M87* agrees with the previous estimate of the $a=(0.9\pm0.5)M$ ~\citep{Tamburini:2019vrf}. 
     For Sgr A*, the EHT infers $\theta_0<50$\textdegree ~, and our analysis independently estimates the inclination angle, consistent with the EHT results. Our results suggest that Sgr A* is spinning rapidly.
    \item To estimate charge, we model M87* and Sgr A* as Kerr-Newman, rotating Bardeen, and Ghosh-Culetu black holes. We report that the charge of M87* is $g\sim0.37M$ and that of Sgr A* is $g\sim0.40M$ considering  $90$\textdegree inclinations. The accuracy of our inferences depends on the accuracy of the EHT measurements. The inferred results provide a good description of the two SMBHs.
    \item Modeling M87* and Sgr A* as rotating Hayward black holes does not yield the correct parameter estimates, which is inconsistent with the results of other models. Therefore, the Hayward model may not be acceptable in describing the two SMBHs. 
    \item If we know the precise values of the supermassive black hole (SMBH) parameters from independent methods, then we can accurately measure the shadow observables.
\end{enumerate}

 Overall, our method is simple and consistent with prior estimations. We show its applicability to SMBHs and find that it estimates any two black hole parameters, intrinsic or extrinsic. Our technique considers the observational uncertainties  contrary to previously proposed methods and gives results subject to the uncertainties in the observational acquirement of the values of EHT observables for an SMBH. We have also obtained estimations for the electric and magnetic charges of the target SMBHs.

Though we have estimated the parameters of only two SMBHs, the EHT and similar experiments may give observational data for other targets in the future, and our technique is easily applicable to all SMBHs. Further, the method is repeatable for a target SMBH over multiple epochs in time and thus may give increasingly accurate estimations with repeated probes and enhanced observational capabilities like the future Earth and space-based VLBI techniques, namely the ngEHT.  It is also extensible to other observables related to the lower order photon rings, which are soon expected to be resolved by the EHT \citep{Ricarte:2022kft}. However, a limitation of our estimation technique is its capability to estimate only two parameters. To overcome this, alternative approaches, such as Bayesian estimation techniques, can estimate more than two parameters of an SMBH. However, the Bayesian estimation techniques often entail significantly higher computational complexity than the method proposed in this paper. Exploring Bayesian estimation techniques for estimating multiple parameters of an SMBH is a promising avenue for future research.

\section*{Acknowledgements}
M.A.\, is supported by a DST-INSPIRE Fellowship, Department of Science and Technology, Government of India. S.G.G. thanks SERB-DST for research project No.~CRG/2021/005771. S.G.G. would like to thank Rahul Kumar Walia for fruitful discussions.

%%%%%%%%%%%%%%%%%%%%%%%%%%%%%%%%%%%%%%%%%%%%%%%%%%
\section*{Data Availability}

We have not generated any original data in the due course of this study, nor has any third-party data been analysed in this article.

%%%%%%%%%%%%%%%%%%%% REFERENCES %%%%%%%%%%%%%%%%%%
% The best way to enter references is to use BibTeX:

\bibliographystyle{mnras}
\bibliography{example}
%%%%%%%%%%%%%%%%%%%%%%%%%%%%%%%%%%%%%%%%%%%%%%%%%%
%%%%%%%%%%%%%%%%%%%%%%%%%%%%%%%%%%%%%%%%%%%%%%%%%%
% Don't change these lines
\bsp	% typesetting comment
\label{lastpage}
\end{document}